\documentclass[graybox]{svmult}

\usepackage{mathptmx}       % selects Times Roman as basic font
\usepackage{helvet}         % selects Helvetica as sans-serif font
\usepackage{courier}        % selects Courier as typewriter font
\usepackage{type1cm}        % activate if the above 3 fonts are
\usepackage{tikz}
\usepackage{soul}
\usepackage{pgfplots}
\usepackage{makeidx}         % allows index generation
\usepackage{amsmath,amssymb,relsize}
\usepackage{graphicx}        % standard LaTeX graphics tool
\usepackage{multicol}        % used for the two-column index
\usepackage[bottom]{footmisc}% places footnotes at page bottom
\newcommand{\R}{\mathbb{R}}

\newcommand{\Hmap}{\vec{H}}
\newcommand{\eAt}{\mathrm{e}^{\A t}}
\newcommand{\Sb}{\vec{S}}
\newcommand{\A}{\vec{A}}
\newcommand{\xnot}{\vec{x}_0}
\newtheorem{assumption}{Assumption}

\usepackage[sort,nocompress]{cite}
\makeindex             % used for the subject index
\begin{document}
	
	\title*{Koopman Performance Analysis of Nonlinear Consensus Networks}
	% Use \titlerunning{Short Title} for an abbreviated version of
	% your contribution title if the original one is too long
	\author{Hossein K. Mousavi, Christoforos Somarakis, Qiyu Sun, and Nader Motee}
	% Use \authorrunning{Short Title} for an abbreviated version of
	% your contribution title if the original one is too long
	\institute{Hossein K. Mousavi \at Dept. of Mechanical Engineering \& Mechanics, Lehigh University, Bethlehem, PA 18015, USA,\\ \email{mousavi@lehigh.edu}
		\and Christoforos Somarakis \at Dept. of Mechanical Engineering \& Mechanics, Lehigh University, Bethlehem, PA 18015, USA, \email{csomarak@lehigh.edu} \and Qiyu Sun \at Dept. of Mathematics, Orlando, FL 32816, USA, \\ \email{qiyu.sun@ucf.edu} \and Nader Motee \at Dept. of Mechanical Engineering \& Mechanics, Lehigh University, Bethlehem, PA 18015, USA, \email{motee@lehigh.edu}}
	%
	% Use the package "url.sty" to avoid
	% problems with special characters
	% used in your e-mail or web address
	%
	\maketitle
	
	\abstract{Spectral decomposition of dynamical systems is a popular methodology to investigate the fundamental qualitative and quantitative properties of {  these} systems and their solutions. In this {chapter}, we consider a class of nonlinear cooperative protocols, which consist of multiple agents that are coupled together via an undirected state-dependent graph. We develop a representation of \text{the} system solution by decomposing the nonlinear system utilizing ideas from the Koopman {operator} theory and its spectral analysis. We use recent results on {the extensions of} the well-known Hartman theorem for hyperbolic systems to establish a connection between the {original} nonlinear {dynamics} and the linearized dynamics in terms of Koopman spectral properties. The expected value of the output energy of the nonlinear protocol, which {is related} to {the} notions of coherence and robustness in dynamical networks, is evaluated and characterized in terms of Koopman eigenvalues, eigenfunctions, and modes. Spectral representation of the performance measure enables us to develop algorithmic methods to assess the performance of this class of nonlinear dynamical networks as a function of their graph topology.  Finally, we propose a scalable computational method for approximation of the components of the Koopman mode decomposition, {which} is necessary to evaluate the systemic performance {measure} of the nonlinear dynamic network. % the Koopman eigenfunctions. 
	}
	

	\section{Introduction}

	The central objective in the theory of networked control systems is to address and analyze the practical challenges in implementations of real-world dynamical networks, in order to develop design algorithms {  with certified convergence properties} %convincing proof certificates
	\cite{beard2001coordination,leonard2001virtual, jadbabaie2003coordination,olfati2004consensus,wang1991navigation,SomBarecc15,dorfler2012synchronization,susuki2012nonlinear}. The application areas, nowadays,  range from multi-robot systems \cite{ali2014review} to social networks \cite{jadbabaie2012non}, power systems \cite{dorfler2013synchronization}, metabolic pathways \cite{van2006parameter, buzi2010quantitative, siami2012existence}, and brain networks \cite{bassett2006small}. One of the inherent unappealing features of these real-world networks is {  the nonlinearity of the} interactions among the subsystems that stem from how subsystems affect each other's dynamics 
	\cite{olfati2004consensus,jadbabaie2004stability,ajorlou2011sufficient,Cucker_Smale_2007,chiang1994bcu,arcak2007passivity}. For example in the natural networks, physical interactions such as fluid field coupling \cite{shilong2001aerodynamic}, coupled biochemical reactions \cite{siami2012existence}, or visual coordination \cite{Cucker_Smale_2007} may result in nonlinear coupling among the subsystems.  
	
	The main focus of the existing body of literature is on stability analysis of nonlinear dynamical networks, where some of {  these works} investigate effects of coupling topologies \cite{dorfler2013synchronization, jadbabaie2004stability}, time-delay \cite{SomBarifac15, papachristodoulou2006synchonization, papachristodoulou2010effects} and exogenous noise \cite{cucker2007flocking}. The common approach to deal with {  the} existing nonlinearities is to study linearized forms of network dynamics. There is a rich number of works devoted to performance and robustness analysis and optimal design of  linear dynamical networks \cite{kim2005maximizing, olshevsky2009convergence, patterson2010leader, bamieh2012coherence,lin2013design,moghaddam2015interior,siami2015performance,siami2016fundamental,mousavi2017spectral,mousavi2017performance, siami2018growing, siami2018network, somarakis2018time, de2016growing,dai2011optimal}. Despite a growing need to analyze and synthesize the nonlinear dynamical networks in non-equilibrium modes of operation, consistent and systematic methods to tackle these problems are sorely missing in the literature. The main reason is that {  the} linear network techniques, which are mainly based on eigendecomposition, cannot be applied to %\st{analysis of} 
	nonlinear 
	%\st{maps}
	{  systems}.   
	%On the contrary, the literature is very poor in results for non-linear versions of distributed coordination algorithms despite the recurrent appearance of non-linear models of co-operative control and synchronization from different fields of study \cite{kuramoto84,Krause00,Cucker_Smale_2007}. The main reason is the absence of a solid underlying theory that can tackle the problem of evaluating systemic performance in a similar vein to the linear systems. 
	Recent advances in  {  analysis of dynamical systems using Koopman operator theory have} %\st{Koopman methods to analyze dynamical systems has}  
	opened up a new venue to study the properties of nonlinear systems in a systematic manner \cite{mauroy2016global,lan2013linearization,budivsic2012applied,Kutz2016,mauroy2017koopman}.  
	
	%Moreover, the results hold promise to extend the current literature in evaluating the performance of non-linear distributed formation control and synchronization algorithms.
	
	In this {chapter}, we build upon concepts and tools from Koopman methodology to assess the performance of a class of nonlinear consensus networks. These networks are defined over an undirected state-dependent interconnection graph topology, where {  the} control input of each agent is equal to a weighted combination of {  the difference} between its own state and its neighbors. The expected value of the output energy of the network is adopted as the performance measure. We obtain a closed-form series representation for this quadratic performance measure and show that the value of performance measure depends on {  the} spectra of the Koopman operator. The idea of spectral characterization of performance measure can be potentially utilized to analyze and design nonlinear networks; we refer to \cite{siami2018growing, siami2018network, somarakis2018time, mousavi2017spectral} for successfulness of this approach {  in the case of} linear dynamical networks. An efficient numerical algorithm is developed to compute {  the} value of the performance measure for a given dynamical network. Several analytical and numerical examples have been provided to highlight the usefulness of our theoretical findings.
	
	\section{{  Preliminaries}}
	
	Consider an autonomous dynamical system {  given by}
	\begin{align}\label{eq:xdot=fx}
	\vec {\dot x}=\vec F(\vec x),
	\end{align}
	with $\vec F(\vec x):\R^n\rightarrow \R^n$ representing a  $C^2$ vector field on $\mathbb R^n$. 
	For the initial condition $\vec{x}_0\in \mathbb R^n$, ${  \vec x(t):=}\vec S(t,\vec {x_0}): \mathbb R_+ \times \mathbb R^n\rightarrow \mathbb R^n$ 
	is the generated flow of \eqref{eq:xdot=fx}, which is assumed to be defined for all $t\geq 0$. We assume that ${\vec F}$ attains a hyperbolic stable fixed point at the origin.
	i.e., $\vec F(\vec 0)=\vec 0$. Moreover, we denote  the Jacobian of $\vec F$ at the fixed point by 
	\begin{align}
	{\vec A}:=\frac{\partial}{\partial \vec x} \vec F|_{\vec x= \vec 0},
	\end{align} 
	which we assume to be Hurwitz; i.e., the eigenvalues of $\vec A$ have strictly negative real parts. The basin of attraction of the origin is an open neighborhood of $\vec 0$ {  with $\Omega \subset \mathbb R^n$ a compact subset of this neighborhood.}  %For the sake of simplicity, we will always assume $S$ compact.
	By definition, ${\vec S}(t,\vec{x}_0)\in \Omega$ for any $\vec{x}_0\in \Omega$ and $t\geq 0$, such that $\vec{S}(t,\vec{x}_0)\rightarrow \vec 0$ as $t\rightarrow +\infty$. 
	{  Let us define the functional space 
		\begin{align}\mathcal F=\bigg\{ f \in C^1(\Omega,\mathbb R) : \sup_{\vec{x}\in \Omega}\big|f(\vec{x})\big|+ \sup_{\vec{x}\in \Omega}\big\|\nabla f(\vec{x})\big\| < \infty  \bigg \}
		\end{align}
		that together with   norm $|f|_{C^1}:= \sup_{\vec{x}\in \Omega}\big|f(\vec{x})\big|+\sup_{\vec{x}\in \Omega}\big\|\nabla f(\vec{x})\big\| $, constitute a Banach space. }
	%{  Let us denote   $\mathcal{F}$ to be a Banach space of functions in $C^1(\Omega,\mathbb{C})$. }
	% $$
	% \st{ {\Psi:=\left \{ f \in C^{1}(\Omega ,\mathbb C)~:~\sup_{\vec x\in \Omega}|f(\vec x)|<\infty{ ,~\sup_{\vec x\in \Omega}|\nabla f(\vec x)|<\infty}\right\}.}}
	% $$
	This will be the space of observable functions on  flow ${\vec S}(\cdot,\vec{x}_0)$.  For fixed $t\geq 0$, the Koopman operator $U^t: {  \mathcal{F}} \rightarrow {  \mathcal{F}} $ associated with \eqref{eq:xdot=fx} is 
	\begin{align}\label{eq: koopman}
	(U^t f)({\vec{x}_0})=f\circ  {\vec S} (t,{\vec{x}_0}).
	\end{align}% \st{Among the basic properties of $U^{t}$ is that it is a linear and bounded operator in $\Psi$.} 
	{  For any fixed $t\geq 0$, it can be shown that $U^{t}$ is linear in $\mathcal{F}$. Furthermore, the collection $\{U^t\}_{t\geq 0}$ constitutes a semigroup known as \textit{the Koopman semigroup} \cite{budivsic2012applied}. In the context of continuous autonomous dynamical systems, \eqref{eq: koopman} is interpreted as the action of semigroup on observable $f\in \mathcal F$.}
	{  The spectrum of operator $U^{t}$ may consist of a discrete, continuous and residual part. The discrete part, also known as point spectrum of $U^t$, is defined as 
		\begin{align}\label{eq:koopmaneig}
		\sigma_p(U^t)=\left \{ \lambda \in \mathbb C \left | ~U^t \phi=e^{\lambda t} \phi,~\text{for some }\phi=\phi_\lambda \in \mathcal{F} \right . \right \}.
		\end{align}
		Throughout this chapter %\st{either} 
		$\big(\lambda,\phi_\lambda\big)$, for $\lambda \in \sigma_p(U^t)$, % or $\big(\mathrm{e}^{\lambda t}, \phi_\lambda \big)$ will be
		is  called the Koopman pair of an eigenvalue with its corresponding eigenfunction.}  
	%Now, let $\big\{ \big(\mathrm{e}^{\lambda t}, \boldsymbol{\phi}_\lambda\big) \big\}_{\lambda\in \sigma_P(U^t)}$ be a countable subset of $\sigma_P(U^t)$. For an observable $f\in \Psi$ that can be represented as 
	% $$f=\sum_{\lambda}v_{\lambda} \phi_{\lambda},$$ we have the following identity  by linearity 
	%\begin{align}\label{eq:koop_eigenfunction}
	%\big(U^{t}f\big) (\vec{x}_0)= \sum_{\lambda} v_\lambda \big(U^{t}\phi_\lambda \big)(\vec{x}_0)=\sum_{\lambda} v_\lambda  \mathrm{e}^{\lambda t} \phi_\lambda(\mathbf{x_0}).
	%\end{align}
	% An eigenfunction of the Koopman operator $\phi(x)$ with an eigenvalue $\lambda \in \mathbb{C}$  is an observable $\phi(x):X\rightarrow \mathbb{C}$ satisfying
	%\phi\circ \Sb(t,x_0)=e^{\lambda t}\phi(x_0),
	%Clearly, one should not expect any member of $\Psi$ to be representable  in terms of the Koopman eigenfunctions. What is more, it is by no means given that $\sigma_{p}(U^t)$ is countable\footnote{Unless $U^{t}$ is at least a self-adjoint operator, it may be the case that $\sigma_p(U^{t})$ is uncountable. See for example in \S II.5.1 in \cite{conway94}.}.
	{  The purpose of this work is to discuss the role of the Koopman operator theory in evaluating quadratic performance measures for a class of nonlinear consensus protocols that enjoy a great interest in the field of networked control systems. More specifically, we leverage a recent extension of the Hartman's theorem for hyperbolic dynamical systems \cite{lan2013linearization} to outline the pivotal role of point spectrum in approximating the output energy of nonlinear distributed cooperative algorithms.}
	
	The rest of the chapter is organized as follows. In \S \ref{sec:koop}, we will apply the extension of the Hartman's theorem in order to investigate the conditions under which one is able to express the flow $\vec{S}(\cdot,\vec{x}_0)$ of \eqref{eq:xdot=fx} in terms of the Koopman pairs, i.e., to write the $i$-th element of $\vec{S}(\cdot,\vec{x}_0)$ as 
	\begin{align*}
	\big[\vec S (t,\mathbf{x_{0}})\big]_i{  \approx} \sum_{\lambda}{c}_{\lambda}^{(i)} \mathrm{e}^{ \lambda t} \phi_{\lambda}(\mathbf{x_0}),~~~~ \text{for every } t\geq 0
	\end{align*}  for some   coefficients $\vec c_\lambda=[c_\lambda^{(1)},\dots,c_\lambda^{(n)}]^T$. Then, each collection    $\{(\lambda,\phi_\lambda,\vec c_\lambda)\}_{\lambda}$   will constitute a Koopman Mode Decomposition (KMD) \cite{budivsic2012applied}. We use an interesting fact about the map created by stacking {  specific} eigenfunctions of the {  Koopman operator} and its inverse map  for 
	dynamical systems 
	with hyperbolic stable fixed points : {  polynomial approximations of the inverse map yields a Koopman Mode Decomposition. }
	
	Based on the results of \S \ref{sec:koop}, we proceed in \S \ref{sect: performance} with the calculation of the performance measures for nonlinear consensus networks. The measures are expressed series form as a function of KMD's. In addition, we discuss a number of special cases where KMD's can be explicitly calculated.
	
	{In Section, \ref{sect: sparseapproximation} we describe a method to come-up with a sparse approximation to the eigenfunctions of the Koopman operator. The method strongly depends on a nearly-optimal fitting technique called Smolyak-Collocation projection. We use the same method to compute the approximate Koopman modes. Using the above developments, we may derive quantitative information about the stability and performance of nonlinear dynamical networks. In fact, inspired by our previous work \cite{mousavi2016koopman}, we look at the performance measure of a class of nonlinear dynamical systems and illustrate how their performance can be assessed using the spectra of the Koopman operator. }

	\section {{  Koopman Mode Decomposition of} System Flows}\label{sec:koop}
	The celebrated theorem of Hartman (stated below for convenience) establishes a crucial connection between {  autonomous dynamical system }\eqref{eq:xdot=fx} and the dynamics of the linearized system around the origin. A moment of reflection, initially mentioned in \cite{lan2013linearization}, can lay the groundwork of bridging the gap between spectral properties of the nonlinear and the linearized system around the fixed point. The aim of the present section is to conduct a rigorous discussion of these exact steps. We begin our analysis with parts adapted from literature to keep the manuscript self-contained. 
	
	\begin{theorem}\label{thm:HG_thm}
		[Hartman's Theorem \cite{perko2013differential} ] Consider   dynamical system \eqref{eq:xdot=fx} with the smoothness assumptions on $\mathbf F$ to hold and the origin to be a hyperbolic fixed point. Then there exists a $C^1$-diffeomorphism $\vec{H}$ of a neighborhood $U$ of the origin on an open set $\Omega'\subset \Omega$ containing the origin such that for each $\xnot \in \Omega'$, there exists is an open interval $I(\mathbf x_0)\subset \mathbb R_+$ containing zero such that for all $\mathbf {x_0} \in U$ and $t\in I(\mathbf {x_0})$ 
		$$
		\Hmap \circ \Sb (t,\mathbf {x_0})=\mathrm{e}^{\A t}\,\vec{H}(\mathbf{x_0}),$$ 
		where $\vec{A}=\frac{\partial}{\partial \mathbf x}\mathbf F|_{\mathbf x=0}$.
	\end{theorem}
	
	\begin{remark} The set $I(\mathbf x_0)$ stands for the maximal interval of existence of the solution of system \eqref{eq:xdot=fx}, that defines   flow $\Sb(t,\mathbf x_0), \text{for any}~t\geq 0$. {  Evidently,  $I(\mathbf x_0)=\mathbb R_+$ for all $\mathbf x_0$ in the basin of attraction $\Omega$. }
	\end{remark}
	
	\noindent The next result extends the theorem of Hartman to hold true over the whole the basin of attraction of the {  fixed point at the} origin.
	
	\begin{theorem}\emph {\cite{lan2013linearization}} 
		If $\mathbf F$ is $C^2$ and  $\A=\frac{\partial}{\partial \mathbf x}\mathbf F|_{\mathbf x=0}$ is Hurwitz, then there exists a diffeomorphism ${\vec{\alpha}}:\Omega \rightarrow \mathbb{R}^{n}$ such that
		\begin{align}\label{eq:mapping}
		\vec{\alpha} \circ\Sb(t,\vec {x_0})=\eAt \, \vec{\alpha} (\vec{x}_0),
		\end{align}
		for all $\vec{ x_0}\in \Omega$ and $t\geq 0$.
	\end{theorem}
	\vspace{1mm} { 
		Next, assuming that $\A$ is diagonalizable, we can write $\A=\vec R {\Lambda} \mathbf{R}^{-1}$ where $\Lambda$ is a diagonal matrix, having diagonal elements with strictly negative real parts.} Let us define  \begin{equation}\label{eq: Hmap}
	\Hmap(\vec x):=\vec R^{-1}\vec{\alpha}(\vec x).
	\end{equation}
	Then, {  one may observe that}
	\begin{align}\label{eq:exp_form}
	\Hmap \left({{\Sb}}(t,\vec{x}_{0})\right)=\vec R^{-1}\eAt \vec R \Hmap(\vec{x}_0)=\mathrm{e}^{{\Lambda} t}{\Hmap}(\vec{{x}_{0}}).
	\end{align}
	Clearly, map $\Hmap:\Omega \rightarrow {  \mathbb{C}^n} $ is a diffeomorphism. {  Hence,    flow of the dynamical system $\Sb(\cdot,\vec{x}_0)$ can be expressed as}
	\begin{align}\label{eq:inverse_mapping}
	\Sb(t,\vec{{x}_{0}})=
	{\Hmap}^{-1}\left(\mathrm{e}^{\Lambda t}{\Hmap}(\vec{{x}_{0}})\right),~~{  \text{ for every~}} t\geq 0 \text{~and~} \vec x_0 \in \Omega.
	\end{align}
	This suggests that knowledge of maps ${\Hmap}$ and ${\Hmap}^{-1}$ %in conjunction with knowledge of $\Lambda$,
	% i.e. the diagonal matrix of the eigenvalues of $A$,
	helps identify the flow of the system. {  In an interesting turn of events, there is an important correlation between Koopman spectrum and the eigenvalues of the Jacobian matrix at the fixed point $\vec A$.}% The next results outlines the necessary bridge between $U^{t}$ and $\Hmap$. 
	\begin{theorem}\label{lem:H(x)}
		Let map ${\Hmap}$ given in \eqref{eq: Hmap}   have component-wise expression \begin{align}\Hmap=\big[H_1,H_2,\dots, H_n\big]^T,
		\end{align}
		for $H_{i}:\Omega \rightarrow {  \mathbb{C}^n}$ and $i =1,\dots,n$. If $\lambda_{i}$ is the $i$-{th}  eigenvalue of $\A$, then $\big({  \lambda_i},\, H_i\big)$ is a pair of Koopman eigenvalue and its corresponding eigenfunction.
	\end{theorem}
	\begin{proof}
		The result immediately follows after comparing  {  the definition of the Koopman eigenfunction in} (\ref{eq:koopmaneig}) with {identity} (\ref{eq:exp_form}).
	\end{proof}
	We take advantage of this connection to provide a {Koopman Mode Decomposition} (KMD) for dynamical systems with a stable hyperbolic fixed point. One may find the general aspects of this decomposition in \cite{budivsic2012applied}.  In this context, Extended Dynamic Mode Decomposition (EDMD) \cite{williams2015data} is a framework with focus on derivation of numerical estimations to Koopman operator {  and KMD}. At first, we make two crucial remarks before coming up with the advertised decomposition.

	\noindent \textit{Polynomial Expansion of  ~$ \Hmap^{-1}$.} 
	Clearly, all elements of $\Hmap^{-1}(\vec{x})=\vec{\alpha}^{-1}(\vec R\, \vec{x})$ are continuous in $\Omega$, hence they map compact sets onto compact sets. Therefore, the domain of definition of $\vec{H}^{-1}$ is compact. 
	
	%{\RC For $\Omega$ compact set, we define
	%\begin{align}
	%\Omega^{-1}:=\{\mathrm{e}^{\Lambda t} \vec H(\vec x_0):~~\text{ for all ~} \vec x_0 \in \Omega,~~ t \in \R_+\}
	%\end{align} \end{comment}
	%Since $\Omega$ is compact and the eigenvalues of $\vec A$ have negative real parts, one observes that set $\Omega^{-1}$ is compact as well. 
	%}
	By virtue of the Stone-Weierstrass Theorem \cite{Cullen1968}  $\vec H^{-1}(\vec x)$ can be {  uniformly $\epsilon$-approximated over the domain of $\vec{H}^{-1}$ by multivariate polynomials.} Therefore, for every $\vec x  \in \text{dom}\,\vec{H}^{-1}$ we can write   %in concise form
	\begin{align}\label{eq:mc_hinv}
	\vec H^{-1}(\vec{x})\,{  \overset{\epsilon}{\approx}}\,\sum_{ { {\gamma\in \Gamma_\epsilon}}} { {\vec{c}}_{{\gamma}}^\epsilon}~ x_1^{j_1}\cdots x_n^{j_n},
	\end{align} 
	where  $\mathbf{\gamma}=[j_1,\dots j_n]^T\in \mathbb{Z}_{+}^n$,  {  $\overset{\epsilon}{\approx}$ implies the approximation with maximal error of $\epsilon$}, and $\vec{c}_{\gamma}^\epsilon=\vec {c}_{j_1,\dots j_n}^\epsilon$ is represented {  using the}  multi-index notation.  {  The index set $\Gamma_\epsilon \subset \mathbb{Z}_+^n$ consists of finite number of indices based on the desired level of accuracy $\epsilon>0$.}
	% There are various rigorous methods for calculation of the coefficients $\{\vec{c}_{\gamma}\}_{\gamma}$. One method is to verify that
	{If map} $\Hmap^{-1}$ is analytic, {  then} it admits a Maclaurin expansion with a positive radius of convergence {  and we can have an infinite series representation (at least in a subset of $\Omega^{-1}$) similar to \eqref{eq:mc_hinv}}.  
	{ 
		\begin{remark} \label{rem:apzero} Not every polynomial approximation of $\vec H^{-1}$ is suitable in this chapter. It is necessary for the right hand-side of \eqref{eq:mc_hinv} to vanish at the origin, as $\vec H^{-1}$ does as well. This property permits a credible polynomial approximation of the output energy of \eqref{eq:xdot=fx} in terms of Koopman modes. Examples of polynomial expansions that can approximate $\vec{H}^{-1}$ under such constraints are interpolation based methods using multi-variate  polynomials of the Bernstein or Chebyshev families, with appropriate scaling of domain of $\vec H^{-1}$ \cite{bernpoly}.  \end{remark}}
	
	%Another method is via identifying $\vec{c}_{\gamma}$ as the Fourier-type projections with respect to orthogonal families of polynomials with several variables. These latter approaches have been systematically investigated in \cite{dunkl2014orthogonal}. In either case, we can assume without loss of generality that a representation of the form \eqref{eq:mc_hinv} is always feasible for any $\vec{x}_0\in \Omega$. 
	
	%We remark however that in numerical explorations in subsequent sections, we restrict to a neighborhood of the fixed point, for computational convenience.
	
	\noindent{\textit{Superposition of Koopman Eigenpairs.}}
	%The second concept is 
	The closedness of the set of eigenfunctions {  under multiplication is an important property} that is stated in the next lemma.
	\begin{lemma}[\cite{budivsic2012applied}] \label{lem:lemma}
		Let $\phi_{1},\phi_{2} \in {  \mathcal{F}}$ with associated eigenvalues $\lambda_1$ and  $\lambda_1$, respectively. Then $\phi_3(\vec x):=\phi_{1}(\vec x)  \phi_2 (\vec x) \in {  \mathcal{F}} $ with associated eigenvalue $\lambda_3=\lambda_1+\lambda_2$%$\mathrm{e}^{\lambda_3 t}=\mathrm{e}^{(\lambda_1+\lambda_2)t}$.
	\end{lemma}
	We are ready now to formulate a KMD-based expression for  flow $\vec{\Sb}(\cdot,\vec{x}_0)$. For its exposition we consider an arbitrary but fixed ordering of the elements of $\mathbb{Z}_+^n$, $\mathbb{Z}_+^n=\{\vec{\gamma}_1,\vec{\gamma}_2,\dots,\vec{\gamma}_i,\dots \}$ where $\vec{\gamma}_i=(j_1,j_2,\dots,j_n)^T$. 
	\begin{proposition} \label{pr:K_decomposition}
		Let  $\vec A=\frac{\partial}{\partial \vec{x}}\vec{F}(\vec{x})|_{\vec{x}=\vec{0}}$ be diagonalizable and Hurwitz {  with eigenvalues $\lambda_1,\dots,\lambda_n$}. Consider map $\vec{H}^{-1}(\vec x)$ with elements given \eqref{eq: Hmap} for every $\vec{x}_0\in \Omega$, {  where $\Omega$ is a compact set}. {  Then, using the approximation \eqref{eq:mc_hinv} for $\vec H^{-1}(\vec x)$,}    flow $\Sb(\cdot,\vec{x}_0)$ of nonlinear system  \eqref{eq:xdot=fx} attains the representation 
		\begin{align*}
		{\Sb}(t,\vec{x}_0)\overset{\epsilon}{\approx}\sum_{i\geq 1} \vec{c}_i \, \mathrm{e}^{{\bar{\lambda}}_i t}\phi_i(\vec{x}_0),~~~\text{ for all } t\geq 0
		\end{align*}
		where  for the ordered vector  $\vec{\gamma}_i={  [j_1,\dots,j_n]^T \in \Gamma_\epsilon} \subset \mathbb{Z}_+^n$, we have 
		\begin{align} \label{eq:lbar}
		\bar{\lambda}_i:=\sum_{k=1}^n j_k  \lambda_{k}~~~\text{and}~~
		~\phi_i(\vec{x}_0):=\prod_{k=1}^n H_{k}^{j_k}(\vec{x}_0).\end{align}
	\end{proposition}
	%{  The uniform precision parameter $\delta_\epsilon$ will go to zero if $\epsilon$ goes to zero. }
	\begin{proof}
		Recall the expression (\ref{eq:inverse_mapping}) that is true for every $\vec{x}_0\in \Omega$. % For $\Hmap(x_0)\in S $, $e^{\Lambda t}\Hmap(x_0)\in S$ for all $t\geq 0$, since $e^{\Lambda t}$ only decreases the magnitude of each component.
		Substituting $\mathrm{e}^{\Lambda t}\Hmap(\xnot)$ into { (finite)}  series representation (\ref{eq:mc_hinv}), we may write the flow of system \eqref{eq:xdot=fx} as
		\begin{align*}
		\vec{\Sb}(t,\vec{x}_0) {  \overset{\epsilon}{\approx}}  \sum_{ {{  \gamma\in {\Gamma_\epsilon}}}} { {\vec{c}}_{{\gamma}}^\epsilon}~ 
		\left (\mathrm{e}^{\lambda_1 t} H_1(\xnot )\right)^{j_1}\dots \left (\mathrm{e}^{\lambda_n t} H_n(\xnot)\right)^{j_n},
		\end{align*} 
		which can be reorganized to obtain
		\begin{align*}
		\vec{\Sb}(t,\vec{x}_0){  \overset{\epsilon}{\approx}}  \sum_{ {{  \gamma\in {\Gamma_\epsilon}}}} { {\vec{c}}_{{\gamma}}^\epsilon}~ 
		\mathrm e^{(j_1\lambda_1+j_2\lambda_2+\dots+j_n\lambda_n)t} H_1^{j_1}(\xnot)\dots H_n^{j_n}(\xnot).
		\end{align*} 
		{  Let us define $\bar{\lambda}_i$ and $\phi_i(\vec x)$ according to \eqref{eq:lbar}.}  Using Lemma \ref{lem:lemma},  we deduce  that $\phi_i(\vec x)$ is an eigenfunction of Koopman operator with eigenvalue  ${\bar{\lambda}_i}$. Rewriting the flow and using the introduced notation gives us the desired representation.
	\end{proof}
	% This is an explicit functions of the spectral components of the linearized system.
	%{  some help in the remark below....}
	%\begin{remark}\label{cor:ev_ef}
	%  The sets of the Koopman eigenvalues and Koopman eigenfunctions of the KMD of the flow $\vec{\Sb(\cdot,\vec{x}_0})$ are 
	%\begin{align}\label{eq:koop_eigvalue}
	%S_{\overline\lambda}:=\left\{\sum_{k=1}^K  \lambda_{i_k}| K\in \mathbb{N},~ i_k =1,\dots,n\right \} ~~\text{and}~~S_{\phi}:=\left\{\prod_{k=1}^K H_{i_k}(\vec{x})| K\in \mathbb{N},~ i_k =1,\dots,n\right\}
	%\end{align}
	%respectively; %, where $\mu_i$ denoting the eigenvalues of $A$. 
	%i.e. the set of  the eigenvalues of the Koopman operator $U^t$ is constructed by the arbitrary {\BR integer weighted sums} of the eigenvalues of $A$, and the set of the eigenfunction with the arbitrary products of those Koopman eigenfunctions with the eigenvalues of $A$, (note that $i_k$ is allowed to repeat). 
	%\end{remark}
	%{  some help in the remark above....}
	
	In fact, {  we  derive the explicit decomposition introduced in Proposition \ref{pr:K_decomposition} by extending the material presented in \cite{mauroy2016global} or \cite{lan2013linearization}.} We will see that this decomposition is a necessary tool for the subsequent analysis. Before that, we recall a useful lemma, that  identifies a partial differential equation {  to associate the Koopman pairs.}
	\begin{lemma} \label{lem:egn_eqn}
		[See \cite{mauroy2016global}]
		{  Consider a pair of Koopman eigenvalue and its corresponding eigenfunction denoted by  $\big(\lambda,\phi_\lambda(\vec{x})\big)$ associated with nonlinear dynamics \eqref{eq:xdot=fx}. The pair satisfies the identity }
		\begin{align}\label{eq:gov_eq}
		{\vec F}(\vec{x})^T \nabla\phi_\lambda(\vec{x})=
		\lambda \phi_\lambda(\vec{x}).
		\end{align}
		% where $\nabla$ denotes the gradient. 
	\end{lemma}

	\section{Performance of Nonlinear Consensus Networks}\label{sect: performance}
	The standard multi-agent setting regards a finite collection of agents labeled as $i=1,2,\dots,n$. The $i$-th agent is characterized by a real-valued state $x_i$. In a {consensus network} with first order dynamics, the agents update their states by communicating with their adjacent (neighboring) agents. Our focus in this work is on the class of dynamic protocols of the form
	\begin{align}\label{eq:agent_dynamics}
	\dot x_i=\sum_{\{i,j\}\in \mathcal{E}} w_{ij}\,(x_j-x_i), 
	\end{align} 
	where $\mathcal{E}$ is the set of edges of the undirected graph of the network whose weights are symmetric and state-dependent in the form of 
	\begin{align}\label{eq:weight_form}
	w_{ij}=w_{ji}=\tilde w_{ij}\,g\left(|x_i-x_j|{  ^2}\right),
	\end{align}
	for  $g:\R_{+}\rightarrow \R_{++}$  a positive coupling function of the graph, and constant $\tilde w_{ij}>0$. We  note that such a state-dependence of the couplings is motivated by a natural assumption: the remote or dissimilar agents less likely interact with each other. For instance, this is the case in the context of social networks, oscillatory networks \cite{jadbabaie2004stability} or biological networks. For this reason function $g$ is usually considered to be monotonically decreasing \cite{Cucker_Smale_2007,SomBarecc15}. By defining the state of the network as $\vec{x}:=[x_1,\dots,x_n]^T \in \R^n$, we may express the collective dynamics of the agents as
	\begin{align}
	\vec{\dot x}=-\mathcal{L}_{\vec{x}}\,\vec{x}
	\end{align}
	where $\mathcal{L}_{\vec{x}}$ is the state-dependent graph Laplacian matrix with coupling weights that vary according to (\ref{eq:weight_form}). For  subsequent analysis we rely on two conditions, stated right below.
	\begin{assumption} \label{assump:graph}The function $g$ is {  analytic} and it satisfies $g(0)=1$.
	\end{assumption}
	\begin{assumption} \label{assump:graph_con}The graph with coupling weights $\{\tilde{w}_{ij}\}_{\{i,j\} \in \mathcal E}$ is  connected.
	\end{assumption} 
	Connectedness implies that there exists a {  linked path} between any two distinct nodes $i$ and $j$ in the graph of the network. A consequence of the latter assumption is that the graph corresponding to $\mathcal L_{\vec{x}}$ remains connected and undirected for all $\vec x\in \R^n$, since $w_{ij}>0$ for every $\{i,j\}\in \mathcal{E}$. The next result  provides a standard sufficient condition for  convergence of {  dynamical network} \eqref{eq:agent_dynamics} to consensus equilibrium.

	\begin{theorem}\label{thm: stability} Let Assumptions \ref{assump:graph}  and \ref{assump:graph_con} hold true. For any initial state $\vec{x}_0\in \R^n$ the long term dynamics satisfy  $$\lim_{t\rightarrow \infty}\vec{\Sb}(t,\vec{x}_0)=\bar x\, \mathbf{1}_n,$$ where the average vector of the network is 
		$ \bar {{ x}}:= { \frac{1}{n}\displaystyle \sum_{i=1}^n x_i(0)}.$ The convergence to consensus occurs exponentially fast, with a rate that depends on   initial state $\vec{x}_0$.
	\end{theorem}
	\begin{proof} At first, observe that $$\max_{i,j = 1,\dots,n}|x_i(t)-x_j(t)|\leq \max_{i,j = 1,\dots,n}|x_i(0)-x_j(0)| \text{ for all } t\geq 0.$$ This is easily verified since for the node $i$ with the maximum initial condition $\max_{i} \dot x_i(t)\leq 0$. Similarly for the node $i$ with the minimum starting value $\min_{i} \dot x_i(t)\geq 0$. The solution $\mathbf x(t,\vec{x}_0)$ remains bounded in $\Omega_0:=[\min_i x_i(0), \max_i x_i(0)]$. Consider the Lyapunov functional $\Lambda(\vec{x})=\dfrac{1}{2}\displaystyle \sum_{i\neq j}|x_i-x_j|^2$. Then for the solution $\mathbf x(t),~t\geq 0$ of \eqref{eq:agent_dynamics}, we have
		\begin{equation*}
		\frac{d}{dt}\Lambda(\vec{x}(t))=\sum_{i\neq j}\big(x_i(t)-x_j(t)\big)\big(\dot{x}_i(t)-\dot{x}_j(t)\big)\leq -\beta(t) \Lambda(\vec{x}(t))
		\end{equation*} where the value of $\beta(t)$ is given by $$\beta(t):=\min_{\mathlarger{s\in[0,t]} \atop \mathlarger{\{i,j\}\in \mathcal E} }w_{ij}\big(\vec{x}(s)\big)\geq \underline{w}\cdot \underline{g}>0$$ for
		$\underline{w}=\displaystyle \min_{i,j=1,\dots,n}\tilde{w}_{ij}$  and $\underline{g}=\, \displaystyle \min_{s_1,s_2\in \Omega_0} g(|s_1-s_2|)>0$ \cite{Mesbahi_Egerstedt_2010}. By virtue of graph connectivity the convergence to the agreement space $x_1=x_2=\dots=x_n$, occurs exponentially fast. Finally, observe that $\frac{1}{n}\displaystyle \sum_{i=1}^n x_i$ is a first integral of motion to conclude about the consensus point.
	\end{proof}
	
	The average of $\vec{x}_0$ is called the consensus {equilibrium} of the network over the state of interest \cite{olfati2004consensus}. The central objective of this work is to evaluate systemic measures of performance that quantify the necessary effort the dynamical system takes to converge to consensus. We aim at leveraging the Koopman framework, developed in the previous section.  The requirement for the implementation of that machinery is to have a hyperbolic and asymptotically stable fixed point. One may notice that %the Jacobian of the consensus network $\mathcal{N}$ with $F(x)=-L(x)x$ at $x=0$ can be evaluated as
	$$\A:=-\mathcal{L}_{\vec 0},$$ with  a smallest eigenvalue in magnitude is $\lambda_1(\A)=0$. Hence, the fixed point at the origin is not hyperbolic. In order to overcome this difficulty we introduce output dynamics vector $\vec{y}$ with elements $y_i:=x_i-\frac{1}{n}\sum_{k=1}^n x_k$, or in matrix form, $\vec{y}=\vec M_n \vec{x}$, where $\vec M_n$ is the the centering matrix given by 
	$$ 
	\vec M_n:=\vec I_n-\vec J_n/n \in \R^{n \times n},
	$$ where $\vec J_n$ is the square matrix of all ones.   The dynamics of $\mathbf y$ constitute the disagreement network associated with \eqref{eq:agent_dynamics}  is defined to pass this obstacle \cite{olfati2004consensus,siami2016fundamental}.
	%\begin{align}\label{eq:dis_dynamics}
	%\vec{\dot x}=-\mathcal {L}_{d}(\vec{x})\, \vec{x},
	%\end{align}
	The disagreement Laplacian matrix is \begin{align*}\mathcal L_d(\vec{x}):=\mathcal L_{\vec{x}}+\frac{\delta}{n} \vec{J}_n \end{align*} for some $\delta>0$. The next stability result is a straightforward corollary of Theorem \ref{thm: stability} and it is stated without proof.
	\begin{corollary}\label{lem:dis} The output dynamics of $\vec{y}=\vec M_n \vec{x}$ of \eqref{eq:agent_dynamics} satisfy
		\begin{equation}\tag{$\mathcal N_d $}\label{eq: disagreementdynamics}
		\vec{\dot y}=-\mathcal L_d(\vec{y})\, \vec{y},
		\end{equation} with $\vec{y}=\vec{0}$ is the a globally exponentially stable hyperbolic fixed point. 
	\end{corollary}
	
	%\begin{definition}
	%For any $\gamma_i=(j_1,j_2,\dots,j_n)^T\in \mathbb{Z}_+^n$, the corresponding restricted triplet  $(\lambda_i,\phi_i(x),C_i)$ is 
	% $$\lambda_i:=\sum_{k=2}^n j_k  \mu_{k}, 
	%~\phi_i(x):=\prod_{k=2}^n H_{k}^{j_k}(x) \text{ and }
	%C_i:= c_{\gamma_i}.$$
	%\end{definition}
	
	%
	%
	%The dynamical system of interest now becomes 
	%\begin{align}\label{eq:consensus_dynamics} \tag{$\mathcal{N}$}
	%\left. \begin{array}{ll}
	%\vec{\dot{x}}=&-\mathcal L_{\vec{x}} \vec{x}\\
	%\vec{y}=&M_n\vec{x}
	%\end{array}. \right .
	%\end{align}
	%For the subsequent analysis, we impose assumptions concerning the function $f$ and  graph with Laplacian $L(x)$. 
	
	%Based on this result, the next lemma points out that the consensus indeed occurs for $\mathcal{N}$. 
	
	%\begin{lemma}\label{lem:stab}{\it\cite{somarakis2015general}}
	%Under Assumptions \ref{assump:graph} and  \ref{assump:graph_con}, for all $x\in \R$, the graph Laplacian $L(x)$ corresponds to a connected undirected weighted graph, $\mathcal{N}$ is  globally stable, and the consensus occurs. Furthermore, once $\mathcal{N}$ is restricted to $\oneperp$, it has a globally asymptotically stable fixed point at the origin.
	%\end{lemma} 
	
	The dynamics of \eqref{eq: disagreementdynamics} satisfy $\vec{y}(t,\vec{y_0})=\vec{M}_n \,\vec{S}(t,\vec{x}_0)$, $t\geq 0$. The energy of the output once weighted with a positive-definite and symmetric matrix $\vec Q$ is 
	$$
	\int_{0}^{\infty} \vec y^T(t,\vec{y_0}) \vec Q~\vec y(t,\vec{y_0}) \,dt.
	$$
	%Consensus is the situation when $x_i(t)\rightarrow \bar x$ for all $1 \leq i\leq n$, or $y(t)\rightarrow 0$. 
	We choose the performance measure as the {mean} energy of the vanishing signal $\vec y$, when the state of the consensus system starts from a random initial condition $\xnot$. %which is related to $\mathcal{L}_2$ norm of $y(t)$. 
	The long term energy of the output signal $\mathbf y$ that converges to zero is equivalent to the energy of the state vector $\vec{x}$ to converge to consensus.
	We take this mean for uncertain initial conditions, by assuming that the initial state is a random variable $\vec{x}_0:\Omega_s \rightarrow \Omega $ from the sample space $\Omega_s$, with some probability measure (e.g. a probability density function or a probability mass function).  % \st{can be taken based on the probability of an initial condition occurring, quantified by a probability density function (pdf)}}. 
	In either case, we define the performance measure as
	% with respect to $(\mathcal{I},g(x))$ is defined as
	\begin{align}\label{eq:per_measure}
	\rho\left(\mathcal{L}\right):=%={E}\left\{\left\|Q^{1/2}y(t)\right \|_{\mathcal{L}_2}^2\right\}=
	\mathbb{E}_{\vec{x}_0}\left\{\int_{0}^{\infty} \vec{S}^T(t,\vec{x}_0) \vec{M}^T_n\vec Q~\vec M_n \vec{S}(t,\vec{x}_0) \,dt\right\}.
	\end{align}
	
	%Before that, we point out that if the initial conditions in the random variable of initial conditions are bounded, then the value of our performance measure would be bounded as well. 
	%
	%\begin{theorem}[Boundedness of the Performance Measure] If initial conditions with respect to which the performance measure (\ref{eq:per_measure}) is defined are almost surely bounded, and also Assumptions \ref{assump:graph} and \ref{assump:graph_con} hold,  then the performance measure is bounded. 
	%\end{theorem}
	%
	%\begin{proof}
	%Due to the assumptions, the first consensus network is stable. Hence, as $t$ increases, $x(t)$ approaches to $x$
	%\end{proof}
	%In this section, we benefit from the KMD that we introduced in Proposition \ref{pr:K_decomposition} to derive a closed-form for the performance measure of interest. 
	
	%In fact, the output dynamics is the disagreement dynamics $\mathcal{N}_d$ once we impose the restriction of $\vec{x}\in \oneperp.$  As the result of this restriction, the Koopman parameters $(e^{\lambda_i t},\phi_i(\vec{x}),\vec{c}_i)$ contributing to the KMD of the output will be restricted as well. These triplets are defined below and will be used in the next theorem.
	The next result establishes an analytical expression for the performance measure of \eqref{eq: disagreementdynamics} that reflects the contributions of the spectra of the linearized graph Laplacian and eigenfunctions of the Koopman operator.
	
	\begin{theorem}\label{thm:perf_measure}
		\emph{(Performance Measure)} Consider the disagreement dynamics \eqref{eq: disagreementdynamics} and the associated flow $\vec{\Sb}(\cdot,\vec{y_0})$ for all initial disagreements $\vec{y_0}$. 
		Then, the performance measure \eqref{eq:per_measure} can be expressed as 
		\begin{align}\label{eq:per_measure_calculated}
		\rho(\mathcal {L})=\sum_{i,j\geq 1}\phi_{ij}c_{ij} \frac{1}{{\bar \lambda}_i+{\bar \lambda}_j},
		\end{align}
		where $\left \{\bar \lambda_i\right \}_{i=1,2,\dots}$ is the sequence of Koopman eigenvalues in the KMD of \eqref{eq: disagreementdynamics}, enumerated by an arbitrary numbering of $\gamma_i =(j_2,\dots,j_n) \in \mathbb{Z}_+^{n-1}$ as 
		$$
		\bar{\lambda}_i:=\sum_{k=2}^n j_k  \lambda_{k}~~~\text{and}~~
		~\phi_i(\vec{x}_0):=\prod_{k=2}^n H_{k}^{j_k}(\vec{x}_0).
		$$
		with $\lambda_2,\dots,\lambda_n$ {  being} the nonzero eigenvalues of $\mathcal{L}_{\vec 0}:=\frac{\partial}{\partial \vec{x}}\mathcal{L}(\vec{x})|_{\vec{x=0}}$. Moreover, $\phi_{ij}:=\mathbb{E}_{\vec{x}_0}\{\phi_i(\vec y)\phi_j(\vec y)\}$ and $c_{ij}:=\vec{c}_i^T Q\, \vec{c}_j$, are computed in terms of Koopman eigenfunctions and modes, respectively.
	\end{theorem}
	
	\begin{proof}
		The disagreement dynamics \eqref{eq: disagreementdynamics} attain a globally exponentially stable hyperbolic origin. In view of Assumptions \ref{assump:graph} and \ref{assump:graph_con}, one can sort the eigenvalues of $-\vec A=\frac{\partial}{\partial \vec{y}}\mathcal L_d(\vec{y})|_{\vec{y}=\vec 0}$ as $\lambda_1<\lambda_2\leq \dots \leq \lambda_n$ such that $\lambda_1=\delta$. We claim that the restriction of $\phi_1(\vec x)=H_1(\vec x)$ to $\mathbf{1}^{\perp}$ is zero, since $\phi_1(\vec x)= \mathbf{1}_n^T \vec x.$
		We substitute $\phi_1(\vec x)$, $\vec F(\vec x)=-\mathcal {L}_d(\vec x)\vec x$, and $\lambda_1=-\delta$ into the left hand side of  (\ref{eq:gov_eq}) to obtain
		\begin{align*}
		\nabla^T\phi_1({\vec x}){\vec F}({\vec x})=\mathbf{1}_n^T (-\mathcal L(\vec x)-\delta \vec J_n/n)\vec x,
		\end{align*}
		which implies that 
		\begin{align*}
		\nabla^T\phi_1({\vec x})\vec {F}({\vec x})=0-\delta\sum_{i=1}^n x_i=- \delta\times \mathbf{1}^T_n \vec x=-\lambda_1 \phi_1.
		\end{align*}
		Therefore, $\phi_1(\vec x)=\mathbf{1}_n^T \vec x$ is in fact a Koopman eigenfunction with eigenvalue $-\delta$. We observe that for any $\vec y\in \mathbf{1}^{\perp}$, it holds that $\phi_1(\vec y)=0$. 
		Considering the restricted dynamics, $H_1(\vec y)=\phi_1(\vec y)=0$. Hence, any Koopman eigenfunction parametrized with $\gamma_i=(j_1,j_2,\dots,j_n)$ with $j_1\geq 1$ is zero, because the corresponding eigenfunction is 
		$$
		\phi_{i}(\vec x)=\prod_{k=1}^n H_{k}^{j_k}(\vec x).
		$$
		Now let a $\Hmap^{-1}$ have the form \eqref{eq:mc_hinv} for $\vec{y}$. We consider a KMD based on Proposition \ref{pr:K_decomposition}. 
		This implies that all terms related to $\lambda_1$ are canceled out of the decomposition. Thus, we can restrict the numbering of summation indices to $\mathbb{Z}_{+}^{n-1}$
		and then write the KMD for $\vec{y}(\cdot ,\vec{y_0})=\vec{M}_n \vec{S}(\cdot,\vec{x}_0)$ as 
		\begin{align}\label{eq:KMD_y}
		\vec{y}(t ,\vec{y_0})=\sum_{i\geq 1}\vec{c}_i \mathrm e^{-{\bar \lambda}_i t}\phi_i(\vec{y_0})
		\end{align}
		where for any multi-index
		$\gamma_i =(j_2,\dots,j_n) \in \mathbb{Z}_+^{n-1}$ inducing  
		$$
		\bar{\lambda}_i:=\sum_{k=2}^n j_k  \lambda_{k}~~~\text{and}~~
		~\phi_i(\vec{x}_0):=\prod_{k=2}^n H_{k}^{j_k}(\vec{x}_0).
		$$
		The integrand of the integral in the performance measure is 
		\begin{align*}
		%x=\sum_{i=1} e^{-{\lambda}_i t}\phi_i(x_0)\mathbf{C}_i\Rightarrow \\
		\vec{y}^T(t ,\vec{y_0})~\vec Q~\vec{y}(t ,\vec{y_0})=\left(\sum_{i\geq 1} \mathrm e^{-{\bar \lambda}_i t}\phi_i(\vec {y_0})\vec{c}_i^T\right)\vec Q
		\left(\sum_{j\geq1} \vec{c}_j \mathrm e^{-{\bar \lambda}_j t}\phi_j(\vec {y_0})\right).
		\end{align*}
		We reorganize this quadratic term as
		\begin{align*}
		\vec{y}^T(t ,\vec{y_0})~\vec Q~\vec{y}(t ,\vec{y_0})=\sum_{i,j\geq 1} \mathrm e^{-({\bar \lambda}_i+{\bar \lambda}_j) t}\phi_i(\vec{ y_0})\phi_j(\vec {y_0})\vec{c}_i^T\vec Q~\vec{c}_j.
		\end{align*}
		%
		%The eigenvalues of $A$ are real, so are the corresponding eigenfunctions. 
		The induced eigenvalues satisfy $\bar \lambda_i=\displaystyle \sum_{k=2}^n j_k  \lambda_{k}>0$, hence, $\bar \lambda_i+\bar \lambda_j>0$ for all $i,j\geq 1$. Integrating over all times yields
		\begin{align*}
		\int_{0}^{\infty} \vec{y}^T(t ,\vec{y_0}) \vec{Q}~\vec{y}(t ,\vec{y_0})~dt=\sum_{i,j\geq 1}\phi_i(\vec{y_0})\phi_j(\vec{y_0}) \frac{\vec{c}_i^T\vec Q \vec{c}_j}{{\bar \lambda}_i+{\bar \lambda}_j}.
		\end{align*} The result follows by virtue of the linearity of the expected value.
	\end{proof}
	\subsection{Analytic Examples}
	The Koopman representation of flows in consensus networks can be derived analytically, for some special cases. In this section, we discuss a few such types of networks in the form of \eqref{eq:agent_dynamics} where the associated Koopman modes \big(subsequently $\rho(\mathcal L)$\big) can be calculated in a closed form.
	\begin{example}[Linear Consensus Network]
		We evaluate the performance measure of a first-order LTI consensus network of order $n$, which has the dynamics
		$$
		\dot {\vec x}=-\mathcal L\, \vec x,
		$$
		for a graph Laplacian $\mathcal L$ that is   state-independent {  (i.e. $g\equiv 1$)} but satisfies Assumption \ref{assump:graph_con}. To use (\ref{eq:per_measure_calculated}), we let $\vec Q=\vec I_n$ and choose the initial conditions such that $$\mathbb{E}_{\vec{x}_0}\left\{ \vec{y_0}\vec{y_0}^T \right\}=\vec I_n.$$  We denote the eigenvalues of $\mathcal L$ as $\lambda_i$ for $i=1,\dots,n$. Based on Lemma \ref{lem:H(x)}, $\lambda_i$ has a Koopman eigenfunction $\phi_{i}(\vec x)=H_i(\vec x)$, that is 
		$$\phi_{i}(\vec x)=\vec v_i^T\vec x,$$ 
		where $\vec v_i$ is the unit eigenvector of $\mathcal L$ corresponding to $\lambda_i$ (see \cite{budivsic2012applied,mauroy2016global}). Let $\vec V=[\vec v_1|\vec v_2|\dots |\vec v_n]$ be the orthonormal matrix of eigenvectors of $\mathcal L$, then for the disagreement dynamics we have $\Hmap(\vec y)=\vec V^T\vec y.$ Since $(\vec V^T)^{-1}=(\vec V^{-1})^{-1}=\vec V$ the inverse of this map is $\Hmap^{-1}(\vec y)=\vec V \vec y.$ This lets us compute the components of the performance measure as follows.
		\begin{align*}
		\phi_{ij}=\mathbb{E}_{\vec{x}_0}\left\{{\phi_i(\vec y)\phi_j(\vec y)}\right\}=\mathbb{E}_{\vec{x}_0}\left\{ \vec v_j^T\vec y \cdot \vec v_i^T\vec y   \right\},
		\end{align*}
		for all $i,j=2,\dots,n$. We rearrange to obtain 
		\begin{align*}\phi_{ij}=\mathbb{E}_{\vec{x}_0}\left\{ \vec v_j^T\vec y\vec y^T \vec v_i   \right\}= \vec v_j^T\mathbb{E}_{\vec{x}_0}\left\{ \vec y\vec y^T\right\}\vec v_i =  \vec v_j^T \vec v_i=\delta_{ij},
		\end{align*}
		since $\mathbb{E}_{\vec{x}_0}\{\vec y\vec y^T\}=\vec I_n$ and $\vec V$ is orthonormal. Obviously, $\Hmap^{-1}(\vec y)=\vec V \vec y$ is a exact polynomial representation, thus
		%Maclaurin series representation, thus
		\begin{align*}
		\vec c_i =
		\left\{
		\begin{array}{ll}
		\vec v_i  & \mbox{if } i=2,\dots,n\\
		\vec {0} & \mbox{if }  i=1
		\end{array}
		\right. ,
		\end{align*}
		%as shown in the proof of Theorem \ref{thm:perf_measure}.
		which allows to compute the coefficients used in the performance measure as
		\begin{align*}
		c_{ij} =\vec c_i^T \vec c_j=
		\left\{
		\begin{array}{ll}
		\delta_{ij}  & \mbox{if } i,j=2,\dots,n \\
		0 & \mbox{if }  i \mbox { or }j=1
		\end{array}
		\right. .
		\end{align*}
		%from which we get
		%\begin{align*}
		%\phi_{ij}c_{ij} =
		%\left\{
		%    \begin{array}{ll}
		%        \delta_{ij}  & \mbox{if } 2 \leq i,j \leq n \\
		%        0 & \mbox{if }  i>n \mbox{ or } j>n
		%    \end{array}
		%\right. .
		%\end{align*}
		We substitute these terms into the result in (\ref{eq:per_measure_calculated}) to find  
		\begin{align}
		\rho(\mathcal{L})=\sum_{i=2}^n{\frac{1}{2\lambda_i}},
		\end{align}
		that is the $\mathcal{H}_2$-norm squared of a first order linear consensus network \cite{siami2016fundamental}.
	\end{example}
	
	It turns out that for the case when we have only two agents, we may be able to compute the eigenfunctions analytically. The next two examples highlight this fact.
	
	\begin{example}\label{ex:alpha_curve} Suppose that the network consists of two agents with dynamics dictated by \eqref{eq:agent_dynamics} and weight functions
		\begin{align}\label{eq:CS}
		w_{ij}=\frac{1}{(1+(x_i-x_j)^2)^\alpha},
		\end{align}
		for some constant $\alpha \in \R_{+}$. Parameter $\alpha$ in (\ref{eq:CS}) defines how localized  the interactions are within the network. As $\alpha$ increases, the agents update their states mainly with respect to their closest neighbors.  In fact, the particular type of link implies that magnitude of interaction between two subsystems becomes weaker as their state becomes more different. For such a consensus network, in the case of two nodes, let $p$ and $q$ denote the states of the agents. Consequently, we can explain the interaction of these two nodes through the dynamics 
		\begin{align*}
		\begin{bmatrix}
		\dot p \\
		\dot q
		\end{bmatrix}=
		\frac{-1}{{(1+ (p-q)^2)^{\alpha}}}\begin{bmatrix}
		1 & -1 \\
		-1 & 1
		\end{bmatrix}
		\begin{bmatrix}
		p \\
		q
		\end{bmatrix}.
		\end{align*}
		Now, we turn into the disagreement dynamics $\mathcal{N}_d$ with $\delta=1$\footnote{Note that choice of $\delta$ is arbitrary.}, whose Jacobian at the origin attains the eigevnalues $\lambda_1=-1$ and $\lambda_2=-2$. Based on Lemma \ref{lem:H(x)}, each eigenvalue corresponds to a Koopman eigenfunction, say $\phi_1(x)$ and $\phi_2(x)$. We restrict the dynamics to $\mathbf{1}_n$, however, for the sake of simplicity, we denote the restricted variables with the same notation  (i.e., $p$ and $q$) . Hence, $[p,q]^T\in \mathbf{1}^{\perp}$; i.e., $p+q=0$. We already know that $$\phi_1(\vec x)=\mathbf{1}_n^T\vec x,$$ whose  restriction to $\mathbf{1}^{\perp}$ is indeed zero. Once $\phi_2(\vec x)$ is restricted to $\mathbf{1}^{\perp}$, we may compute it in an explicit fashion using (\ref{eq:gov_eq}), that is
		\begin{align*}
		\begin{bmatrix}
		\partial \phi_2 / \partial p \\
		\partial \phi_2 / \partial q
		\end{bmatrix}^T
		\frac{-1}{{(1+(p-q)^2)^\alpha }}
		\begin{bmatrix}
		q-p \\
		p-q
		\end{bmatrix}=-2\phi_2.
		\end{align*}
		We let $z:=p-q$ and use the chain rule to obtain
		$$\frac{\partial \phi_2 }{ \partial p}=\frac{\partial \phi_2 }{ \partial z}\mbox{ and } \frac{\partial \phi_2 }{ \partial q}=-\frac{\partial \phi_2 }{ \partial z}.$$ Noting that the only free variable is now $z$, we can change the partial derivatives with respect to $z$. The resulting scalar ordinary differential equation is 
		\begin{align*}
		\frac{2z}{(1+z^2)^\alpha}\frac{d\phi_2}{dz}=2\phi_2,
		\end{align*}
		which {  together with $\phi_2(0)=0$ implies that}
		%is equivalent to the following indefinite integral equation
		%\begin{align*}
		%\int \frac{d\phi_2}{\phi_2}=\int \frac{(1+z^2)^\alpha}{z}dz.
		%\end{align*}
		%We add $+1$ and $-1$ into the numerator of the integrand in the right hand side and integrate to get
		%\begin{align*}
		%\ln|\phi_2|=\ln|z| + \int \frac{(1+z^2)^\alpha-1}{z} dz, %\Rightarrow \\
		%&\Rightarrow \phi_2=\pm z \exp\left({ \int \frac{(1+z^2)^\alpha-1}{z} dz}\right),
		%\end{align*}
		%that is equivalent to  
		\begin{align*}
		\phi_2=\pm z \exp\left({ \int \frac{(1+z^2)^\alpha-1}{z} dz}\right).
		\end{align*}
		Without loss of generality, we choose to work with the plus sign. Using the binomial series we write the numerator of the integrand as
		$$h(z):=(1+z^2)^{\alpha}-1=\alpha z^2+\frac{(\alpha)(\alpha-1)z^4}{2!}+\dots,$$
		for all $|z|<1$. We integrate the series to get
		\begin{align*}
		\int \frac{h(z)}{z}dz=%\frac{\alpha z^2}{2}+\frac{\alpha (\alpha-1) z^4}{4\times 2!}+\dots= \\
		\sum_{n=1}^{\infty} \frac{\alpha(\alpha-1)\dots(\alpha-n+1) z^{2n}}{2n \times n!}
		\end{align*}
		which completes the evaluation of $\phi_2(z)$ as 
		\begin{align*}
		\phi_2(z)=z\exp\left(\sum_{n=1}^{\infty} \frac{\alpha(\alpha-1)\dots(\alpha-n+1) z^{2n}}{2n \times n!}\right),
		\end{align*}
		that is a convergent series for $|z|<1$, since $\exp(.)$ is analytic everywhere. The identity $p+q=0$ implies $z=2p$, thus
		\begin{align*}
		\phi_2(p)=2p\exp\left(\sum_{n=1}^{\infty} \frac{\alpha(\alpha-1)\dots(\alpha-n+1) 2^{2n-1}p^{2n}}{n \times n!}\right).
		\end{align*}
		Thus, the component-wise description of  $\Hmap(p,q)$ is
		$$
		\Hmap(p,q)=\Hmap(p)=
		\begin{bmatrix}
		\phi_2(p) &
		0
		\end{bmatrix}^T.
		$$
		The definition of inverse of a map implies 
		$$
		\Hmap^{-1}(\Hmap(p,q))=\begin{bmatrix}
		p &
		q
		\end{bmatrix}^T
		$$
		so $H_1^{-1}(p,q)$ and $H_2^{-1}(p,q)$ are simply the inverse functions of $\phi_2(p)$ and $-\phi_2(p)$, respecively. These functions are locally analytic around  the origin based  on Lagrange inversion theorem \cite{abramowitz1966handbook}. Futhermore, one can calculate their coefficients in terms of Bell polynomials \cite{charalambides2002enumerative}. 
		%The same theorem lets us compute the coefficients of the Maclaurin series of $\psi(p)$. 
		We assume that the initial value of $p$ to come from a discrete random variable that takes the values from $\{0,0.1,0,2,0.3,0.4\}$ with the uniform probability distribution.   Because the mean of each initial condition must be zero, the initial value of $q$ will be $-p$. We compute the value of $\rho(\mathcal{L})$ for $\vec Q=\vec I_n$ using the $17$ order Maclaurin series of both eigenfunctions and the inverse map $\Hmap^{-1}(p)$. Because the value of the performance measure may be computed from the numerical integration of the trajectory as well, we may compare the results of the analytic approximations to the KMD and the real value of performance measure. These two value for a range of $\alpha\in [0,0.4]$ have been illustrated in Fig. (\ref{fig:vs_alpha}, where they are in good numerical agreement. The relative error is observed to increase from zero in the case that $\alpha=0$ to less than $0.13 \%$ for $\alpha=0.4$. 
		
		%\begin{figure}[!t]
		%\centering
		%    \centering
		%    \newlength\figureheight 
		%    \newlength\figurewidth 
		%    \setlength\figureheight{4cm} 
		%    \setlength\figurewidth{7cm}
		%    \input{plots/alpha_plot_edited.tikz}
		%    \caption{The performance measure of the network of two agents in Example \ref{ex:alpha_curve} with the decaying parameter $\alpha$.  }
		%    \label{fig:vs_alpha}
		%\label{fig:sim}
		%\end{figure}
		
		\begin{figure}[!t]
			\centering
			\centering
			\includegraphics[width=8cm]{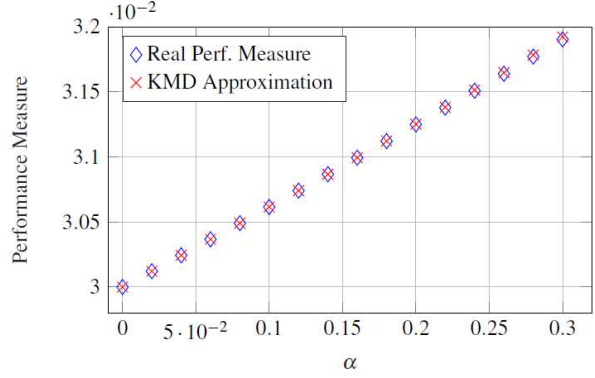}
			\caption{The performance measure of the network of two agents in Example \ref{ex:alpha_curve} with the decaying parameter $\alpha$.  }
			\label{fig:vs_alpha}
		\end{figure}

		%\begin{figure}[!t]
		%\centering
		%\includegraphics[width=2.4in]{images/localization_plot}
		%\caption{\small The CS-Model weight function for different values of $\alpha$ depicting the localization of each coupling function.}
		%\label{fig:sim_3}
		%\end{figure}
		%\end{remark}
	\end{example}
	
	\begin{remark}
		There is a limitation on the magnitude of the admissible initial conditions for the analysis conducted in the previous example. However, notice that this does not imply that it is a linear analysis since for any value of $\alpha$, the linearization matrix of $\mathcal{N}_d$ (with $k=1$) in Example \ref{ex:alpha_curve} is the following matrix. 
		$$
		\A=\begin{bmatrix}
		1 & -1 \\
		-1 & 1
		\end{bmatrix}+\vec J_n.
		$$
		This means that the value of the performance measure computed using the linearized system for each value of decaying parameter $\alpha$ is only one value.
	\end{remark}

\begin{example} \label{ex:K_curve}The dynamics of oscillators have been observed to be closely related to the consensus dynamics. Kuramoto suggested a model of biological oscillation, in which each oscillator was connected to the other one; i.e., the topology was a complete graph. Instead the interactions can be limited over a certain graph \cite{jadbabaie2004stability}, so the dynamics of agent $i$ can be represented as
	\begin{align}\label{eq:KR}
	\dot x_i=\omega_i+\sum_{\{i,j\}\in \mathcal{E}} w_{ij}(x_j-x_i),
	\end{align}
	where $\omega_i$ is the natural frequency and the coupling weight is 
	\begin{align}
	w_{ij}=K\frac{\sin(x_i-x_j)}{x_i-x_j}.
	\end{align}
	When the agents are identical { (i.e., when $\omega_i=\omega$ for some $\omega$)}, the change of variable $x_i \rightarrow x_i-\omega t$ induces a nonlinear consensus network that is
	\begin{align}\label{eq:KR_id}
	\dot x_i=\sum_{\{i,j\}\in \mathcal{E}} w_{ij}(x_j-x_i).
	\end{align}
	
	\noindent We proceed with a procedure for computation of performance measure similar to one introduced in Example \ref{ex:alpha_curve}. For two identical oscillators, with phases $\theta$ and $\gamma$, the dynamics are
	\begin{align*}
	\frac{d}{dt}\begin{bmatrix}
	\theta \\
	\gamma
	\end{bmatrix}= K \,
	\begin{bmatrix}
	\sin(\gamma-\theta) \\
	\sin(\theta-\gamma)
	\end{bmatrix},
	\end{align*}
	Setting $k=1$, the disagreement Jacobian has eigenvalues $\lambda_1=-1$ and $\lambda_2=-2K$, with eigenfunctions $\phi_1$ and $\phi_2$, respectively. Again we only need the restriction of $\phi_2$ to $\mathbf{1}^{\perp}$. The equation (\ref{eq:gov_eq}) for these dynamics becomes
	\begin{align*}
	\begin{bmatrix}
	\partial \phi_2 / \partial \theta \\
	\partial \phi_2 / \partial \gamma
	\end{bmatrix}^T
	{K \sin(\theta-\gamma)}{}
	\begin{bmatrix}
	1 \\
	-1
	\end{bmatrix}=-2K\phi_2.
	\end{align*}
	The new variable $z:=\theta-\gamma$ creates a single ordinary differential equation that is
	\begin{align*}
	\frac{2\sin(z)}{d\phi_2}{dz}=2\phi_2 \Rightarrow \frac{d\phi_2}{\phi_2}=\frac{1}{\sin(z)}dz,
	\end{align*}
	which is integrated and manipulated to  get
	\begin{align*}
	\phi_2=\pm \exp\left(-\ln(\cot(z/2))\right)
	%\pm\frac{1}{\cot(z/2)}
	=\pm \tan(z/2),
	\end{align*}
	where we arbitrarily  choose $+$. The eigenfunction $\phi_2$ is locally analytic around the origin for $|z|<\pi$. Restricting to $\mathbf{1}^{\perp}$, $\theta+\gamma=0$, so $z=2\theta$, hence
	\begin{align*}
	\phi_2(\theta,\gamma)=\phi_2(\theta)=\tan(\theta),
	\end{align*}
	which helps write the components of $\Hmap(\theta,\gamma)$ as $\Hmap(\theta,\gamma)=\begin{bmatrix}
	\phi_2 & 0
	\end{bmatrix}^T$. First component of $\Hmap^{-1}(\theta)$ satisfies
	$
	H_1^{-1}(\phi_2(\theta),0)=\theta.
	$
	Hence, $H_1^{-1}(\theta,\gamma)=\arctan(\theta),$ 
	which is again locally analytic for $|\theta|<1$ around zero. 
	
	We sample initial conditions from a uniform discrete random variable of $63$ equally distributed initial conditions in $\theta \in [0,\pi/5]$ (and  $\gamma=-\theta$) with equal distance of $0.01$. We set $\vec Q=\vec I_n$ and use $17^{\text{th}}$ order Maclaurin series of eigenfunctions and $\Hmap^{-1}(\theta)$ to assess $\rho(\mathcal L)$. As shown in Fig. \ref{fig:kplot}, we alter $K\in [0.2,4]$ and take a look at the values of the performance measure, once compared with the exact value of $\rho(\mathcal L)$ (computed with the numerical integration of the trajectories). The numerical agreement in this experiment can be measured by the relative error of the performance using the KMD approximation, which was about $3\times 10^{-4} \%$ in the worst-case.  %for the case that had the worst error. 
	
	\begin{figure}[!t]
		\centering
		%    \centering
		%    \newlength\figureheight 
		%    \newlength\figurewidth 
		%    \setlength\figureheight{4cm} 
		%    \setlength\figurewidth{7cm}
		%    \input{plots/K_plot_edited.tikz}
		\includegraphics[width=8cm]{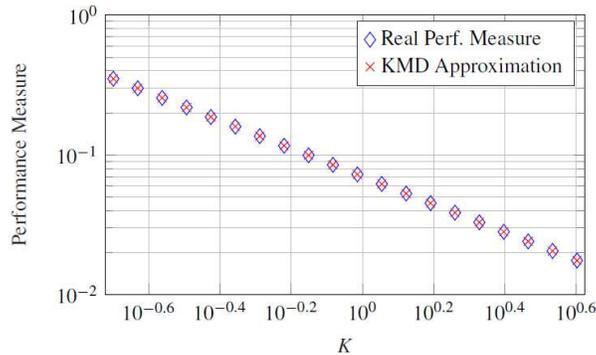}
		\caption{The performance measure of the Kuramoto model of two agents in Example \ref{ex:K_curve} with the parameter $K$.  }
		\label{fig:kplot}
	\end{figure}
	
\end{example}

\section{Sparse Polynomial Approximations}\label{sect: sparseapproximation}

We have observed that any polynomial approximation to the inverse map of eigenfunctions $\Hmap^{-1}(\vec x)$ will result in the Koopman Mode Decompositions. In this section, first, we detail a general sparse approximation technique for multivariate interpolation. Then, we demonstrate how we can use this technique for the map of Koopman eigenfunctions $\Hmap(\vec x)$ as well as the inverse $\Hmap^{-1}(\vec x)$.

\subsection{Smolyak-Collocation Method} 

To introduce the notion of sparsity for the approximation of both Koopman eigenfunctions and Koopman Mode Decomposition, we may use sparse functional approximation methods. The idea is that instead of searching for the approximant in the whole space of polynomials, the search is carried out over a nearly optimal sparse basis, called \emph{Smolyak basis}. The output of the method would be a polynomial: the weighted sum of the tensor product of Chebyshev polynomials that are in the basis. Naturally, we choose the coefficients of the polynomial with respect to some error criterion. One way of doing so is collocation, where we enforce the approximant to (perhaps approximately) satisfy the governing equation of the problem at the given points of a grid, called Smolyak Sparse Grid. To describe this method, we need few basic tools (consult \cite{barthelmann2000high} and \cite{malin2011solving} for more details). 

\begin{definition}[Chebyshev Polynomials] The sequence of the (scalar) Chebyshev polynomials of first kind $\{T_{i}(x)\}_{i=1,2,\dots}$ are initialized with
	$T_1( x)=1$ and $T_2(x)=x$ and recursively defined as follows. 
	\begin{align}
	T_{i+1}(x)=2xT_i(x)-T_{i-1}(x),\mbox{ for   } i=2,3,\dots
	\end{align}
	Similarly, the Chebyshev polynomials of second kind $\{U_{i}(x)\}_{i=1,2,\dots}$ start with $U_1(x)=1$ and $U_2(x)=2x$, and then iteratively
	\begin{align}
	U_{i+1}(x)=2xU_i(x)-U_{i-1}(x),\mbox{ for   } i=2,3,\dots
	\end{align}
	
\end{definition}
We define the integer function $m(i): \mathbb{N}\rightarrow \mathbb{N}$ with $m(1)=1$ and for $i=2,3,\dots$, it is evaluated according to
\begin{align} m(i):=2^{i-1}+1. \end{align}
We also define the  sequence of sets $\{\mathcal{G}^{i}\}_{i=1,2,\dots}$ wherein, $\mathcal{G}^{1}=\{0\}$, and for $i=2,3,\dots$, it holds that $\mathcal{G}^i=\{\zeta_1,\dots,\zeta_i\}\subset [-1,1]$, that is the set of the extrema of the Chebyshev polynomials with the components given by
\begin{align}
\zeta_j:=-\cos\left(\frac{\pi(j-1)}{i-1}\right), \mbox{ for all } j\in\{1,\dots,i\}.
\end{align}
In the next definition, we use the multi-index notation $\mathbf{i}=(i_1,\dots,i_n) {  \in \mathbb{N}^n}$ {  inducing $|\mathbf i|:=\sum_{k=1}^n i_k$. }
\begin{definition}[Smolyak Sparse Grid] The Smolyak Sparse grid of $[-1,1]^n$ is a union of the Cartesian products of the form 
	\begin{align}\mathcal{H}^{n,\mu}:=\bigcup_{|\mathbf{i}|=n+\mu}\left (\mathcal{G}^{m(i_1)}\times \dots \times \mathcal{G}^{m(i_n)}\right ),\end{align}
	where positive integer $\mu \in \mathbb{N}$ is   the order of the grid\footnote{  One can show that grids of higher order include all grids of lower order; i.e., $\mathcal{H}^{n,\mu} \subset \mathcal{H}^{n,\mu+1}$. }. 
\end{definition}

\begin{definition}
	[Smolyak Approximant Polynomial] The Smolyak approximant to a function $f:[-1,1]^n\rightarrow \R$ is given by 
	\begin{align}\label{eq:smo}
	\hat f^{n,\mu}(\vec x):= \mathlarger{\sum}_{q \leq |\mathbf{i}|\leq n+\mu}(-1)^{n+\mu-|\mathbf{i}|} {n-1 \choose n+\mu- |\mathbf{i}|} \mathrm{p}^{\mathbf{i}}{ (\vec x)},
	\end{align}
	with { $q =\max(n, \mu+1)
		$} the tensor product polynomials for each multi-index $\mathbf{i}=(i_1,\dots,i_n)$ defined as
	\begin{align}\label{eq:pol}
	\mathrm{p}^{\mathbf{i}}(\vec x):=\sum_{l_1=1}^{m(i_1)}\dots \sum_{l_n=1}^{m(i_n)} \theta_{l_1,\dots,l_n} T_{l_1}(x_1)\dots T_{l_n}(x_n),
	\end{align}
	where $\theta_{l_1,\dots,l_n}$ the  coefficients  {  that are to be determined}.
\end{definition}

To find the optimal vector of coefficients of ${ \Theta}=\mathrm{vec}(\theta_{l_1,\dots,l_n})\in \R^m$ for approximation of some function $f$ that is $C^k([-1,1]^n)$, an error objective should be defined and minimized. One way is to consider the error function 
$E(f,\Theta):C^k([-1,1]^n)\times \R^n \rightarrow \R_+$ to be
\begin{align*}
E(f,\Theta):=&\underset{[-1,1]^n}{\int}\left |f(\vec x)-f^{n,\mu}(\vec x)\right |^2 \prod_{\vec y\in \mathcal{H}^{n,\mu}}\delta(\vec x-\vec y)~\mathrm{d}\vec{x} \\
& = \sum_{\vec y \in \mathcal{H}^{n,\mu}} \left |f(\vec y)-f^{n,\mu}(\vec y)\right |^2,
\end{align*}  
{  where $\delta$ is the Dirac's delta function.}
This metric is certainly minimized (i.e., $E(f,\Theta)=0$) if $\Theta$ is chosen such that 
\begin{align}\label{eq:col}
f(\vec x)=f^{n,\mu}(\vec x) \mbox{ for all } \vec x\in  \mathcal{H}^{n,\mu}.
\end{align}
This procedure is called collocation. A pivotal property of the overall method is that size of  vector of coefficients $\Theta$ and the number of 
interpolation points is equal; i.e., $|\mathcal{H}^{n,\mu}|=|\Theta|=M.$  
Therefore,  enforcing   equalities (\ref{eq:col}) constitutes of searching for the solution to $M$ equations involving $M$ unknown entries of $\Theta$. Once we 
evaluate the coefficients with the described scheme, the following error bound  would hold, wherein the used functional norm $\|.\|:C^k\left({[-1,1]^n}\right)\rightarrow \R_{+}$ is defined as 
\begin{align}
\|f\|=\max\left \{ {\left \|D^i f\right \|_{\infty}: ~i = 1, \dots, k}\right \}.
\end{align}
\begin{theorem}[{  Theorem 2 in \cite{barthelmann2000high}}]\label{thm:error_sm}
	Suppose that   function $f(\vec x):[-1,1]^n \rightarrow \R$ is $C^k\left({[-1,1]^n}\right)$, together with a Smolyak approximant $\hat f^{n,\mu}(\vec x)$ that interpolates $f$ on $\mathcal{H}^{n,\mu}$ {  with $|\mathcal{H}^{n,\mu}|=M$}. Then, {  for some positive constant $c_{n,k}$},  the error of the approximation is bounded according to 
	\begin{align} \label{eq:errorbound}
	\left \|f-\hat f^{n,\mu}\right \| \leq c_{n,k}M^{-k}(\log M)^{(k+2)(n+1)+1}.
	\end{align}
\end{theorem}

%Because for a fixed $q$, the number of grid points satisfies
% $M\leq c_q n^q,$ \cite{malin2011solving}, 
%we can infer that 
%error bound \eqref{eq:errorbound} also implies that 
%\begin{align}
%\left \|f-\hat f^{n,\mu}\right \| \leq c_{n,k}c_q n^{-qk}(\log c_q+q \log n)^{(k+2)(n+1)+1}.
%\end{align}
%

Each multi-index $\mathbf{i}$ in  (\ref{eq:smo}) induces a number of tensor product polynomials that are summed together as in (\ref{eq:pol}). We gather the indices of all these tensor product
polynomials in a set $\mathbf{L}^{n,\mu}$; i.e.,
\begin{align}
\mathbf{L}^{n,\mu}:=\bigcup_{q \leq |\mathbf{i}|\leq n+\mu} \{(l_1,\dots,l_n): l_j\leq m(i_j) \}. 
\end{align}
One can show that at the end of the day, the approximant constructed in (\ref{eq:smo}) using the polynomials (\ref{eq:pol}) boils down to the following simple 
representation
\begin{align}\label{eq:simpleform}
\hat f^{n,\mu}(\vec x)=%\sum_{i=1}^{M} 
\sum_{\mathbf{l}_i=(l_1^i,\dots,l_n^i)\in \mathbf{L}^{n,\mu}} \Theta_{i} \mathrm{T}_{i}(\vec x)=\sum_{i=1}^{M}  \Theta_{i} \mathrm{T}_{i}(\vec x), 
\end{align}
where $M=|\mathcal{H}^{n,\mu}|=|\mathbf{L}^{n,\mu}|$, and for $\mathbf{l}_i=(l_1^i,\dots,l_n^i)\in \mathbf{L}^{n,\mu}$, 
the coefficients ${\Theta}_{i}$  and polynomial terms are given by 
\begin{align}
& {\Theta}_{i}:= \theta_{l_1^i,\dots,l_n^i}, \\
& {\mathrm{T}}_{i}(\vec x):= T_{l_1^i}(x_1)\dots T_{l_n^i}(x_n).
\end{align}
{  To compute the partial derivatives of approximation, we use the definitions of the Chebyshev polynomials to define }
\begin{align}
{\mathrm{T}}_{i}^j(\vec x):=\left\{
\begin{array}{ll} {\mathrm{T}}_{i}\cdot \dfrac{U_{l_j^i}(x_j) }{T_{l_j^i}(x_j)}
& ~~~~~l_j^i=2,\dots,n \vspace{2mm} \\ 
~~~~~~~~~~0 & ~~~~~l_j^i=1
\end{array}
\right.,
\end{align}
This lets us write the partial derivatives of $\hat f^{n,\mu}$ {  in the compact form  }
\begin{align}\label{eq:partial}
\frac{\partial \hat f^{n,\mu}(\vec x)}{\partial x_j}=\sum_{i=1}^M l_j^i{\Theta}_{i} \mathrm{T}_{i}^j(\vec x).
\end{align} 

\subsection{Sparse Approximation to Eigenfunctions}

We denote the approximation to Koopman eigenfunction $\phi(\vec x)$ by $\hat \phi(\vec x)$. Substituting \eqref{eq:simpleform} and \eqref{eq:partial} into (\ref{eq:gov_eq}), we get the (approximate) equality 
$$
\sum_{j=1}^n \sum_{i=1}^{M} l_j^i{\Theta}_{i} \mathrm{T}_{i}^j(\vec x) F_j(\vec x)\approx \lambda \sum_{i=1}{M}{\Theta}_{i}\mathrm{T}_{i}(\vec x).
$$
We change the order of summations  to further obtain 
\begin{align}\label{eq:atheta}
\sum_{i=1}^{M}  \left ( \sum_{j=1}^n \left (l_j^i \mathrm{T}_i^j(\vec x) F_j(\vec x) \right )-\lambda \mathrm{T}_i(\vec x) \right ) {\Theta}_i \approx 0.
\end{align}
We define and denote the vector of coefficients $\Theta \in \R^M$ by 
$
\Theta:=[\Theta_1,\dots,\Theta_M]^T. 
$ For a point in the grid $\vec x^k\in \mathcal{H}^{n,\mu}$, we may write the left hand side of (\ref{eq:atheta}) as 
$$
\mathcal{A}_{k} {\Theta}= \left [\mathcal{A}_{ki}\right ]_{i=1,\dots,n}  {\Theta},
$$
where the entries of   row vector $\mathcal{A}_k \in \R^{1 \times M}$ can be computed from 
$$
\mathcal{A}_{ki}:= \sum_{j=1}^n \left (l_j^i \mathrm{T}_i^j(\vec x^k) F_j(\vec x^k)\right )-\lambda \mathrm{T}_{i}(\vec x^k), \text{ ~for all } i=1,\dots,M.
$$
Repeating this for $M$ points in the Smolyak grid, the stacked left hand side of all equations becomes $\mathcal{A} \Theta$, where $\mathcal{A}\in \R^{M\times M}$ is given by %consists of rows equal to $\mathcal{A}_k$ as 
\begin{align}
\mathcal{A}:=[\mathcal{A}_1^T,\dots,\mathcal{A}_M^T]^T.
\end{align} 
Ideally, $\mathcal{A}\Theta$ should be zero for an eigenfunction, however, if it is not possible, we would like to minimize an error function, which we choose to be 
\begin{align}\label{eq:J}
J(\Theta)=\|\mathcal{A}\Theta\|_2^2.
\end{align} 

{  Now, because $\vec A$ may have repeated eigenvalues, we add a constraint that lets us derive multiple eigenfunctions   corresponding to one eigenvalue.} 
Consider the Koopman eigenfunction $\phi(\vec x)$ with Koopman eigenvalue $\lambda$, which is the eigenvalue of the linearization matrix with a left eigenvector $\vec R=[\vec r_1,\dots,\vec r_n] \in \R^{n\times n}$. We omit the index of the eigenvalues and eigenvectors in the following developments for simplicity and consider $\lambda$ to be associated with the eigenvector $\vec r \in \R^n$.  We can show that
with the fixed point at the origin,
$$
\nabla \phi (\vec x)|_{\vec x=\vec 0}=\vec r. 
$$
Translating this for the approximant, for each $j=1,\dots,n$ we have
$$
\frac{\partial \hat \phi(\vec x)}{\partial x_j}|_{\vec x=\vec 0}=\sum_{i=1}^M l_j^i{\Theta}_{i} \mathrm{T}_{i}^j(\vec 0)=\mathcal{B}_j\Theta= r_j,
$$ 
where the row vector $\mathcal{B}_j \in \R^{1\times M}$ has the components 
$$
\mathcal{B}_{ji}=l_j^i\mathrm{T}_{i}^j(\vec 0), \text{~~for all }i=1,\dots,M. 
$$
The matrix form of this equality becomes 
\begin{align}\label{eq:wconstraint}
\mathcal{B} \Theta=\vec r,
\end{align}
where the matrix $\mathcal{B}\in \R^{n \times M}$ is the result of stacking row vectors as
\begin{align}\label{eq:mathb}
\mathcal{B}=\left [\mathcal{B}_1^T,\dots,\mathcal{B}_n^T\right ]^T.
\end{align} 
{  Recall that our approximation requires $\hat \phi(\vec 0)=0$ to provide exponential convergence for the performance measure integrals (see Remark \ref{rem:apzero}). This condition can be translated to single scalar equality} 
\begin{align}\label{eq:thetaconstraint}
\mathcal{C}\Theta=0,
\end{align} 
where $\mathcal{C} \in \R^{1 \times M}$ is the row vector with elements 
$$
\mathcal{C}_{i}:=\mathrm{T}_{i}(\vec 0) \text{~~for all }i=1,\dots,M. 
$$

Now, {  we} would like to minimize the error function defined by \eqref{eq:J}, while constraints \eqref{eq:wconstraint} and \eqref{eq:thetaconstraint} are satisfied. We define the optimization problem 
\begin{align}\label{eq:opt_eigen}
&\underset{\Theta \in \R^m }{\mathrm{minimize }}~~ \|\mathcal{A}\Theta\|_2^2, \\
&\mathrm{subject~to }~ \begin{bmatrix}
\mathcal{B} \\ 
\mathcal{C}
\end{bmatrix}\Theta=\begin{bmatrix}
\vec r \\ 
0
\end{bmatrix}. \notag
\end{align}
This program is equivalent to a Semi-Definite Program (SDP) and can be solved using the conventional convex programming such as CVX \cite{grant2010cvx}.  

\subsection{Sparse Approximation to Koopman Mode Decomposition}

In the previous subsection, we illustrated a way to find approximations to the Koopman eigenfunctions. Thus, the components of the map $\Hmap(\vec x)$ can be approximated.  Here, following a similar approach, we seek approximations to the components of its inverse map $\Hmap^{-1}(\vec x)$. The very natural equation for component $H_j^{-1}(\vec x)$  is
\begin{align}\label{eq:inv}
H_j^{-1}(\Hmap(\vec x))=x_j, ~~\text{ for all } j =1,\dots,n. 
\end{align}
We only have an approximation to $\Hmap(\vec x)$, namely $\hat \Hmap(\vec x)$, and we need approximations to $\hat \Hmap^{-1}(\vec x)$, namely $\hat \Hmap^{-1}(\vec x)$. Hence, we consider the approximate equality  
\begin{align}\label{eq:inv_hat}
\hat H_j^{-1}(\hat \Hmap(\vec x))\approx  x_j, ~~\text{ for all } j =1,\dots,n. 
\end{align}
Again, following the spirit of the collocation method, for each point of the grid $\vec x^k \in \mathcal{H}^{n,\mu},$ we  enforce this equation to hold. Suppose that we have found the series approximation to each components of $\Hmap(\vec x)$, denoted by $\hat \Hmap(\vec x)$. Moreover, we define% $H(x^k)$
\begin{align}
\vec z^k:= \hat \Hmap(\vec x^k).
\end{align} 
Then, we consider a Smolyak series representation for this function as
\begin{align}\label{eq:ghat}
\hat g_j^{n,\mu}(\vec z^k)=\sum_{i=1}^{M} {\Phi}_{i}^j \mathrm{T}_{i}(\vec z^k). 
\end{align}
Similar to essence of the method that we discussed in the previous subsection, we define   vector of coefficients $\Phi_1^j \in \R^M$ to be 
$$\Phi^j:=\left [\Phi_1^j,\dots,\Phi_M^j \right ].$$ 
Inserting \eqref{eq:ghat} into   (\ref{eq:inv}), we get 
\begin{align}\label{eq:997}
\mathcal{D}_{k} {\Phi}^j= \left [\mathcal{D}_{ki}\right ]_{i=1,\dots,M}  {\Phi}^j\approx x_j^k,
\end{align}
where the components of   row vector $\mathcal{D}_{k} \in \R^{1 \times M}$ are
\begin{align}
\mathcal{D}_{ki}:= \mathrm{T}_{i}(\vec z^k).
\end{align}
Concatenating these vectors and the right hand side scalars for each point  in the grid (i.e., $M$ points), we may write these equations as
\begin{align}\label{eq:hinv_opt}
\mathcal{D}{\Phi}^j\approx \vec X_j,
\end{align}
where matrix $\mathcal{D} \in \R^{M \times M}$ and  vector $\vec X_j \in \R^M$ are  given by 
\begin{align}
& \mathcal{D}:=\left [\mathcal{D}_1^T, \dots,\mathcal{D}_M^T \right ]^T,\\
& \vec X_j:=\left [x_j^1,\dots,x_j^M \right ]^T,
\end{align}
respectively. Again  one hopes that (\ref{eq:hinv_opt}) holds with a minimal error for each $j=1,\dots,n$. Therefore, we  define the optimization problem
\begin{align}\label{eq:opt_inv_map}
\underset{\Phi^j \in \R^M}{\mathrm{minimize }}~~ \left \|\mathcal{D}\Phi^j-\vec X_j \right \|_2^2.
\end{align}
Note that   matrix $\mathcal{D}$ is deliberately denoted without index $j$, because it is the same matrix for the optimization problem for each
component $H_j^{-1}(\vec x)$. The solution to this least-squares optimization problem is  given by 
$$
\Phi^j=\mathcal{D}^{\dagger} \vec X_j ~~\text{ for all } j =1,\dots,n.  
$$
We should repeat this for each component of $\Hmap^{-1}(\vec x)$. Putting the results in a matrix gives us 
$$
\Phi:=\left [\Phi^1,\dots,\Phi^n \right ].
$$
Because the value of  matrix $\mathcal{D}$ is shared between $M$ optimization problems defined by (\ref{eq:opt_inv_map}), by inspection, we find that
\begin{align}\label{eq:overall_inv}
\Phi=\mathcal{D}^{\dagger} \vec X^T,
\end{align}
where $\vec X\in \R^{n\times M}$ is the matrix containing the vector of all  grid points.  {  Now, we have a polynomial approximation to $\vec H^{-1}$, which would give us a Koopman Mode Decomposition. }

%\subsection{Additional Sparsity: Adding $\ell_1$ Penalizers} \label{subsec:l1}
%
%The sparsity introduced by application of Smolyak grids is  \emph{intrinsic}, as the chosen basis for 
%functional approximation, compared to the basis of all polynomials, is itself sparse. On the other hand,
%if we need or prefer sparser approximations, we may leverage the existing proxies to the sparsity
%such as $\ell_1$ penalization [x]. In the context of finding approximations to the eigenfunctions, this approach 
%has been recently employed as well [x].  The optimization problem (\ref{eq:opt_eigen}) using this technique
%can be modified as
%\begin{align}\label{eq:opt_eigen_ell}
%&\underset{\Theta}{\mathrm{minimize }}~ \|\mathcal{A}\Theta\|_2+\gamma \|\Phi \|_1, \\
%&\mathrm{subject~to }~ \begin{pmatrix}
%\mathcal{B} \\ 
%\mathcal{C}
%\end{pmatrix}\Theta=\begin{pmatrix}
%w \\ 
%0
%\end{pmatrix}, \notag
%\end{align}
%while the optimization (\ref{eq:overall_inv}) defined for finding and approximation to the inverse map $\Hmap^{-1}(x)$ would be
%\begin{align}\label{eq:opt_inv_map_ell}
%\underset{\Phi}{\mathrm{minimize }}~ \left \|\mathcal{D}\Phi-X^T \right \|_2+\gamma \| \Phi \|_1.
%\end{align}

\subsection{Numerical Examples}

In this section, we show that we may be able to effectively estimate the performance measure of nonlinear consensus networks with more than two subsystems.  Note that in all cases, the real performance measure is calculated from the numerical solution of the network output followed by numerical integration. 

\begin{example}[Complete Graphs] Using the described numerical approximation method, we estimate the performance measure for the nonlinear consensus network with exponentially decaying weights defined in \eqref{eq:CS}. The corresponding linearized 
	graph Laplacian corresponds to the undirected unweighted complete graph; i.e.
	$$
	\vec A=-\mathcal{L}_{\mathcal{K}_n}=\vec J_n/n-\vec I_n.
	$$ 
	We evaluate the performance measure from the KMD approximation based on the numerical integration of the solutions. The initial conditions are
	uniformly sampled random initial conditions from $[-1,1]^n$. The numerical values for data using Koopman approach have been obtained by implementation of the suggested numerical method with and the results are shown in Fig. \ref{fig:comp_graphs.tikz}.  In this example, the error in the evaluated performance measure using our numerical method is less than $ 2\%$. 
	
	\begin{figure}[!t]
		\centering
		%    \centering
		%    \newlength\figureheight 
		%    \newlength\figurewidth 
		%    \setlength\figureheight{4cm} 
		%    \setlength\figurewidth{7cm}
		%    \input{plots/comp_graphs.tikz}
		\includegraphics[width=8cm]{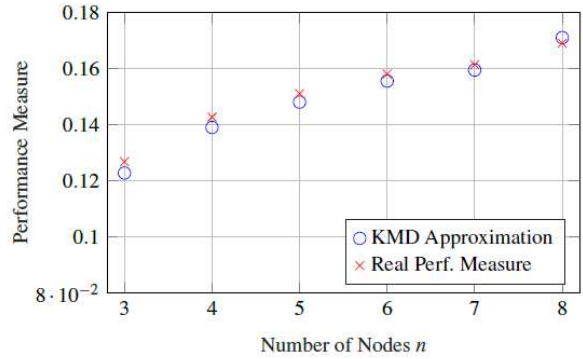}
		\caption{The performance measure of nonlinear consensus network with $\alpha=0.25$ and the graph at the linearized Laplacian of complete graph.  }
		\label{fig:comp_graphs.tikz}
	\end{figure}
\end{example} 

\begin{figure}[!t]
	\centering
	\centering
	\includegraphics[width=8cm]{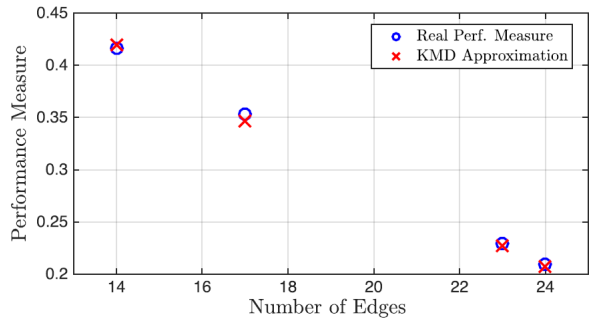}
	\caption{The performance measure of nonlinear consensus network with $n=8$, $\alpha=0.25$ and  random graphs with different number of edges  }
	\label{fig:random}
\end{figure}

\begin{example}[Random Graphs] We fix $n=8$ and create Erd\H os-R\'enyi graphs with different edges probabilities (and consequently, different edge numbers).  Then, we consider again the exponentially decaying weights given by \eqref{eq:CS}. The performance measures from Monte-Carlo simulations as well as the formula (using the method discussed in the previous sections) are also evaluated and illustrated in Fig. \ref{fig:random}. The error of approximation, in this case, is less than $1.4 \%$. 
\end{example}

\subsection{Comparison to Extended Dynamic Mode Decomposition}

The numerical method explained in this section is related to the notion of Extended Dynamic Mode Decomposition (EDMD) \cite{williams2015data}, which has been a promising procedure for extracting information about the Koopman spectrum of the dynamical system.  In EDMD, to find a KMD for the flow of the dynamical system, one should first assume a rich enough {dictionary} of basis functions such that hopefully, the Koopman eigenfunctions lie in their span. Then, using the snapshots from the trajectory, one may find a truncated approximation to the Koopman operator and finite number of approximations to the Koopman eigenfunctions and their corresponding eigenvalues. Then, the solution to the dynamical system is approximated as a truncated KMD using those eigenvalues and eigenfunctions. 

In the current settings, we know what are the eigenvalues and eigenfunctions that are required for representation of the flow of the nonlinear system. Hence, we do not need the first step of the EDMD for the computation of the (approximate) eigenvalues and eigenfunctions. In fact, we {build} the dictionary that one needs for EDMD based on the principal eigenfunctions in the map $\Hmap(\vec x)$. 

On the other hand, the second step in both methods are connected in the spirit:    In our approach, we find an approximation to map $\Hmap^{-1}(\vec x)$ using identity
$$
\Hmap^{-1}\left (\Hmap(\vec x)\right )=\vec x. 
$$
While in EDMD, the identity observable (i.e., $\mathbf f(\vec x)\equiv \vec x$) has to be represented in terms of the basis functions in the dictionary. {  Then one is allowed to write down a KMD for the system dynamics as explained before. 
}
\section{Conclusion and Discussion}{ 
	Koopman mode decomposition approaches hold promise for performance analysis and synthesis of nonlinear dynamical systems, that are of interest in various disciplines of engineering and control. The vital connection between the eigenspectrum of linearized dynamics and Koopman operator provides a closed form evaluation of the first moment of energy integral of the solutions, in terms of Koopman eigenvalues, eigenfunctions, and modes. The numerical approximation of KMD components is implemented by a scalable computational algorithm using sparse Smolyak grid with certifiable accuracy. Future directions include, but are not limited to the following directions: investigation and analysis of various performance metrics in nonlinear systems as extensions of linear control systems \cite{siami2016fundamental}. Another research line regards dynamical systems with higher order integrators as well as a systemic performance-based network synthesis for optimal interactions among interconnected entities in the face of uncertain initial conditions or other structural parameters. }

\section*{Acknowledgements} 

The authors would like to thanks Prof. Alex Mauroy, Prof. Igor Mezic, {and the anonymous reviewer} for fruitful discussions and comments {that enhanced the quality of our manuscript.} 
\bibliography{mybib}

\end{document}